\documentclass[prc,aps,twocolumn,showpacs]{revtex4}
\usepackage{graphicx}
\begin{document}

\title{Ratios of Elastic Scattering of Pions from\\ ${\mathbf{^{3}H}}$
and ${\mathbf{{^{3}}He}}$ }
\author{W.~J.~Briscoe}
\affiliation{Center for Nuclear Studies, Department of Physics,\\
The George Washington University, Washington DC 20052}
\author{%
 B.~L.~Berman}
\affiliation{Center for Nuclear Studies, Department of Physics,\\
The George Washington University, Washington DC 20052}
\author{%
R.~W.~C.~Carter} \altaffiliation{Present address: 1627 Camden Ave. \# 307, Los Angeles,
CA 90025}
\affiliation{Center for Nuclear Studies, Department of Physics,\\
The George Washington University, Washington DC 20052}
\author{%
K.~S.~Dhuga}
\affiliation{Center for Nuclear Studies, Department of Physics,\\
The George Washington University, Washington DC 20052}
\author{%
S.~K.~Matthews} \altaffiliation{Present address: Oregon Hearing Research Center, Oregon
Health Science University Portland, OR 97201-3098}
\affiliation{Center for Nuclear Studies, Department of Physics,\\
The George Washington University, Washington DC 20052}
\author{%
 N-J.~Nicholas}
\altaffiliation{Present address: Advanced Nuclear Technology Group (NIS-6), MS J562,
Nonproliferation and International Security Division, LANL, Los Alamos, NM 87545}

\affiliation{Center for Nuclear Studies, Department of Physics,\\
The George Washington University, Washington DC 20052}

\author{S.~J.~Greene}
\address{Los Alamos National Laboratory, Los Alamos NM 87545\\
}
\author{B.~M.~K.~Nefkens and J.~W.~Price}
\address{Department of Physics, University of California at Los Angeles,
Los Angeles, CA 90095\\
}
\author{L.~D.~Isenhower and M.~E.~Sadler}
\address{Department of Physics, Abilene Christian University,
Abilene TX 79699\\
}
\author{I.~\v{S}laus and I.~Supek}
\address{Rudjer Bo\v{s}kovi\'{c} Institute, Zagreb, Croatia}
\date{\today}

\begin{abstract}
We have measured the elastic-scattering ratios of normalized yields for charged pions
from ${^{3}}H$ and ${^{3}}He$ in the backward hemisphere. At 180~MeV, we completed the
angular distribution begun with our earlier measurements, adding six data points in the
angular range of $119^{\circ }$ to $169^{\circ }$ in the $\pi $ -nucleus center of
mass. We also measured an excitation function with data points at 142, 180, 220, and
256~MeV incident pion energy at the largest achievable angle for each energy $-$
between $160^{\circ}$ and $170^{\circ }$ in the $\pi$-nucleus center of mass. This
excitation function corresponds to the energies of our forward-hemisphere studies. The
data, taken as a whole, show an apparent role reversal of the two charge-symmetric
ratios $r_{{1}}$ and $r_{{2}}$ in the backward hemisphere. Also, for data $\ge
100^\circ$ we observe a strong dependence on the four-momentum transfer squared $-t$
for all of the ratios regardless of pion energy or scattering angle, and we find that
the superratio $R$ data match very well with calculations based on the
forward-hemisphere data that predicts the value of the difference between the
even-nucleon radii of ${^{3}}H$ and ${^{3}}He$. Comparisons are also made with recent
calculations incorporating different wave functions and double scattering models.
\end{abstract}


\maketitle
%%\twocolumn

\section{Background}

The first nucleus to exhibit `real' nuclear properties is the three-body nucleus. There
are only two stable $A=3$ nuclei, ${^{3}}H$ and ${^{3}}He$; neither member of this
isospin doublet has any known excited state. The ground state of ${^{3}}H$ has one
proton and two spin-paired neutrons, while that of ${^{3}}He$ has one neutron and two
spin-paired protons. We refer to the neutrons in ${^{3}}H$ and the protons in
${^{3}}He$ as the `even' nucleons, and to their {\em rms} radii as the even radii for
${^{3}}H$ and ${^{3}}He$, respectively. Similarly, the proton in ${^{3}}H$ and the
neutron in ${^{3}}He$ are called `odd' nucleons and their {\em rms} radii shall be
referred to as odd radii~\cite{Schiff64}. Charge symmetry requires that, in the absence
of Coulomb effects and mass differences, these nuclei be identical. This is because
${^{3}}H$ and ${^{3}}He$ are mirror nuclei and are related by a $180^{\circ }$ rotation
in isospin space, or to put it more simply, one nucleus changes into the other by
interchange of neutrons and protons.

There are four interactions available between the two charged pions and ${^{3}}H$ and
${^{3}}He$ target nuclei. These are grouped conveniently in `isospin' pairs $\pi
^{+3}H$ and $\pi ^{-3}He$, $\pi ^{-3}H$ and $\pi ^{+3}He$ which in turn are related to
each other by the interchange of protons with neutrons and $\pi ^{+}$ with $\pi ^{-}$.
\emph{In the absence of the Coulomb force}, if the strong interaction is charge
symmetric, scattering cross sections for the members of each of these pairs should be
equal.

If we make the assumption that an incident pion interacts with a single target nucleon,
then we can look at these interactions in terms of $\pi $-nucleon scattering
amplitudes. At 180~MeV (near the peak of the $\Delta _{33}$ resonance) the $\pi^+p$
($\pi^-n$) scattering amplitude is about three times as large as the $\pi^+n$
($\pi^-p$.) Consequently, the elastic cross sections for $\pi^{+}p$ and $\pi ^{-}n$ are
nine times as large as those for $\pi ^{+}n$ and $\pi ^{-}p$. The $\pi $-N scattering
amplitudes can be broken down into \textit{spin-flip} and \textit{non-spin-flip}
components. For a fixed target nucleon and pions of 180~MeV, the \textit{spin-flip}
amplitude has a sine dependence on the scattering angle; the \textit{non-spin-flip}
amplitude has a cosine dependence. Since no excited states are available, single
\textit{spin-flip} scattering from one of the even nucleons would leave both nucleons
with the same spin and is thus forbidden by the Pauli principle. (See Ref.~\cite{Pil91}
for further discussion on this point.)

However, at the large scattering angles $\geq 100^\circ$ for which our most recent data
were obtained, kinematic considerations require that one consider double- or
multiple-scattering effects. This would open the door to double spin-flip scattering in
the even nucleons which would not violate the Pauli principle. The description of the
whole picture is discussed in the accompanying article of Kudryavtsev \textit{et
al.}~\cite{SK02}. Pion-nuclear amplitudes, discussed in~\cite{SK02}, are a
superposition of a single- and double-scattering of pions from the nucleons in the
nucleus with Fermi motion taken into account. Principal sources of the violation of
charge symmetry in~\cite{SK02} are (i) the Coulomb interaction between the charged
pions and the nuclei (the external Coulomb effect), (ii) n-p doublet and the mass
splitting of the different charged states of the $\Delta_{33}$-isobar, and (iii) the
difference between the wave functions of $^{3}H$ and $^{3}He$ due to the additional
Coulomb repulsion between the two protons in the $^{3}He$ nucleus (the internal Coulomb
effect.)

Several ratios of scattering cross sections have been defined in conjunction with our
earlier measurements~\cite{Pil91,Nef90,WJB93,KSD96} in order to highlight certain
features of the $\pi - ^{3}\mathrm{A}$ interaction.

The first two are the \textit{charge-symmetric} ratios
\begin{equation}
r_{1}=\frac{d\sigma (\pi ^{+3}H)}{d\sigma (\pi ^{-3}He)}
\end{equation}
and
\begin{equation}
r_{2}=\frac{d\sigma (\pi ^{-3}H)}{d\sigma (\pi ^{+3}He)}.
\end{equation}
The numerator and denominator of each pair are related by an exchange of protons with
neutrons and $\pi ^{+}$ with $\pi ^{-}$. Therefore, the strong interactions in the
denominator and numerator are equal, and these ratios should be identically one at all
energies and four-momentum transfers squared. However, since the form factor of
${^{3}}He$ is decreased owing to the Coulomb repulsion between its protons, the cross
section in the denominator is reduced and we expect that the ratios $r_{1}$ and $r_{2}$
will be somewhat greater than one~\cite{KKL87,BS89}. In $r_{1}$, scattering from the
odd nucleon dominates in the delta energy region, and so the scattering is a mixture of
spin-flip and non-spin-flip. In $r_{2}$, scattering from the even nucleon dominates,
and so spin-flip scattering is suppressed.

The next ratio is $R$, the ``superratio.'' It is defined as the product of $r_{1}$ and
$r_{2}$,
\begin{equation}
R=r_{1}\cdot r_{2}.
\end{equation}
Since $r_{1}$ and $r_{2}$ are charge symmetric, so is $R$.

Finally, we define
\begin{equation}
\rho ^{+}=\frac{d\sigma (\pi ^{+3}H)}{d\sigma (\pi ^{+3}He)}
\end{equation}
and
\begin{equation}
\rho ^{-}=\frac{d\sigma (\pi ^{-3}H)}{d\sigma (\pi ^{-3}He)}.
\end{equation}
These ratios are not charge symmetric, but, as shall be seen
below, several experimental uncertainties cancel when calculating
$\rho ^{+}$ and $\rho ^{-} $, and they can be used to derive the
charge-symmetric $R,$ where
\begin{equation}
R=\rho^{+}\cdot\rho^{-}.
\end{equation}

In Section~\ref{sec:expt}, we briefly discuss the analysis of our data at energies
spanning the $\Delta _{33}$ resonance. The resulting charge-symmetric ratios $r_{1}$
and $r_{2}$, non charge-symmetric ratios $\rho^+$ and $\rho^-$, and Superratio $R$ are
presented in Section~\ref{sec:ratio}. In Section~\ref{sec:90deg}, we discuss the
situation for angles greater than $100^{\circ}$. Finally, we summarize our findings in
Section~ \ref{sec:concl}.

\section{Experiment}
\label{sec:expt}

The experimental details have been given by Matthews \textit{et al.}~\cite{Mat95} .
Here, we briefly discuss the analysis of the data and the relevant experimental
parameters for the determination of the scattering ratios.

\begin{table*}
\caption{ Measured values of the ratios from this experiment. The quoted uncertainties
are statistical only.
%The angle is the $\pi-^{3}\mathrm{A}$ center-of-mass angle.
\newline
}
\begin{ruledtabular}
\label{tab:values}%
\begin{tabular}{cccccccc}
$T_{\pi}(MeV)$ &$\theta$(deg) &$-t(fm^{-2})$ & $\rho^{+}$ & $\rho^{-}$ & $r_1$ & $r_2$
& $R$ \\ \hline 142 & 160.0 &5.0 & 0.811(0.024) & 1.347(0.065)& 1.10(0.06) &
0.99(0.06)&1.09(0.06) \\
142 & 163.6 &5.0 & 0.757(0.027) & 1.401(0.086)& 1.08(0.05) & 1.01(0.06)&1.09(0.06) \\
\hline 180 & 119.4 &5.2 & 0.85 (0.02) & 1.28 (0.04) & 1.07(0.04) &
1.02(0.04)&1.09(0.04) \\
180 & 129.8 &5.7 & 0.87 (0.02) & 1.38 (0.04) & 1.13(0.04) &
1.06(0.04)&1.20(0.05)\\
180 & 139.1 &6.1 & 0.75 (0.03) & 1.56 (0.08) & 1.15(0.06) &
1.01(0.07)&1.16(0.08) \\
180 & 148.3 &6.4 & 0.55 (0.02) & 2.05 (0.09) & 1.09(0.05) &
1.03(0.06)&1.13(0.06) \\
180 & 157.4 &6.7 & 0.44 (0.01) & 2.62 (0.11) & 1.08(0.05) &
1.06(0.06)&1.14(0.06) \\
180 & 169.2 &6.9 & 0.38 (0.01) & 3.06 (0.15) & 1.12(0.06) & 1.05(0.06)&1.18(0.06) \\
\hline 220 & 169.3 &8.9 & 0.408(0.035) & 2.86 (0.18) & 1.21(0.11) &
0.97(0.07)&1.17(0.09) \\ \hline 256 & 169.4 &10.9& 0.478(0.035) & 2.59 (0.37) &
1.25(0.20) & 0.99(0.16)&1.24(0.20)
\end{tabular}
\end{ruledtabular}
\end{table*}

%%\onecolumn
% === PSFIG 1 ======================================
\begin{figure*}
\centering{
\includegraphics[height=0.8\textwidth,angle=90]{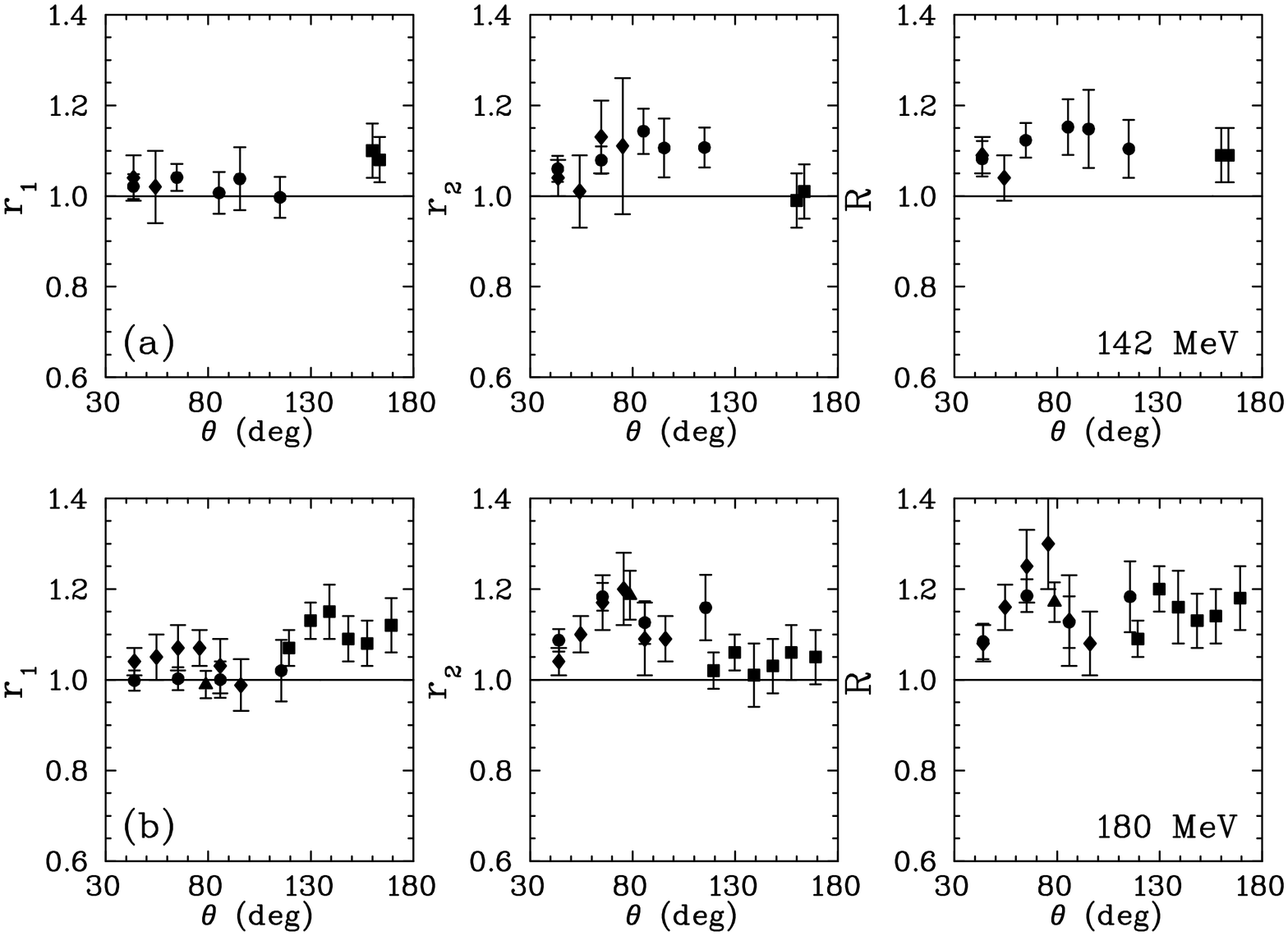}

\vspace{2.5ex}

\includegraphics[height=0.8\textwidth, angle=90]{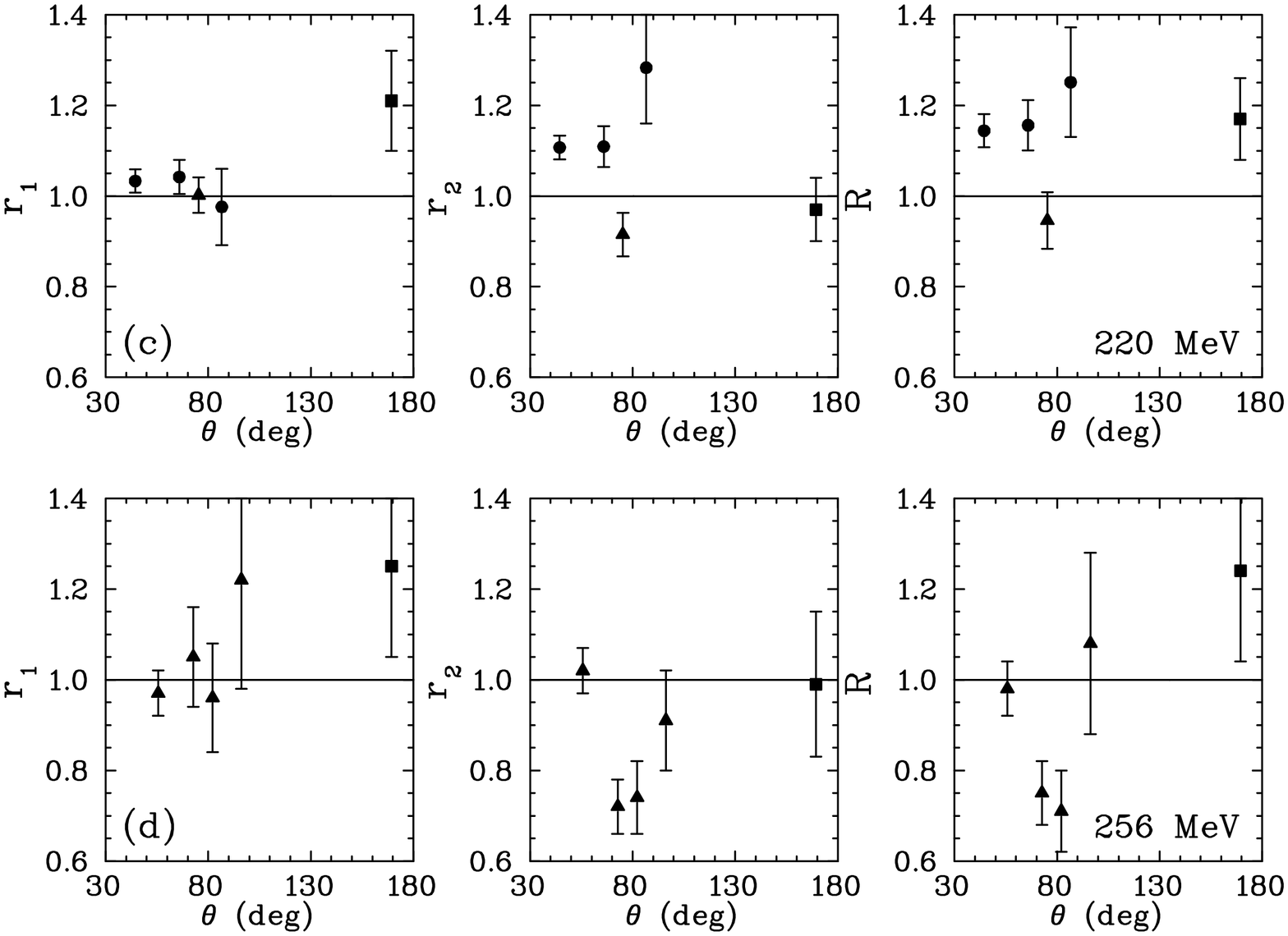}
}
\caption{The ratios $r_1$, $r_2$, and the Superratio $R$
         for ${\pi ^{\pm}}{^3H/^3He}$ elastic scattering
         plotted versus the center-of-mass angle in the
         $\pi-^{3}\mathrm{A}$ system for various incident pion
         kinetic energies: (a) T$_{\pi}$ = 142~MeV,
         (b) 180~MeV, (c) 220~MeV, and (d) 256~MeV.
         Experimental data are from~ \protect\cite{Nef90}
         (diamonds), \protect\cite{Pil91} (circles),
         \protect\cite{KSD96} (triangles), and
         this experiment (squares). \label{fig:r1r2}}
\end{figure*}
%%\twocolumn

The ratios $r_{1}$\ and $r_{2}$ are extracted as

\begin{equation}
r_{1}=\frac{Y(\pi ^{+ 3}H)}{Y(\pi ^{+2}H)}\cdot \frac{Y(\pi ^{-2}H)}{Y(\pi
^{-3}He)}\cdot \frac{d\sigma (\pi^{+2}H)}{d\sigma (\pi ^{-2}H)}\cdot
\frac{N_{^{3}He}}{N_{^{3}H}}
\end{equation}

and
\begin{equation}
r_{2}=\frac{Y(\pi ^{-3}H)}{Y(\pi ^{-2}H)}\cdot \frac{Y(\pi ^{+2}H)}{Y(\pi
^{+3}He)}\cdot \frac{d\sigma (\pi ^{-2}H)}{d\sigma(\pi ^{+2}H)}\cdot
\frac{N_{^{3}He}}{N_{^{3}H}},
\end{equation}
where $Y(\pi ^{\pm }$ $^{n}\mathrm{A})$ refers to the scattering yield, and $N_{^{3}H}$
and $N_{^{3}He}$ are the number of scattering centers per unit volume in the ${^{3}}H$
and ${^{3}}He$ samples, respectively. These latter were measured very accurately $-$
see Ref.~ \cite{Mat95} for details. Elastic scattering yields from $^{2}{H}$ are used
to scale the other yields to the known $\pi ^{\pm 2}H$ cross sections. Writing the
ratio $d\sigma (\pi ^{+ 2}H)/d\sigma (\pi ^{-2}H)$ as $D$, and defining $Y_{N}$ as the
yield per target nucleon $Y(\pi ^{\pm}$ ${^{n}\mathrm{A})}/N_{A}$, we have
\begin{equation}
r_{1}=\frac{Y_{N}(\pi ^{+3}H)\cdot Y(\pi ^{-2}H)\cdot D}{Y_{N}(\pi
^{-3}He)\cdot
Y(\pi ^{+2}H)}
\end{equation}
and similarly
\begin{equation}
r_{2}=\frac{Y_{N}(\pi ^{-3}H)\cdot Y(\pi ^{+2}H)}{Y_{N}(\pi
^{+3}He)\cdot Y(\pi ^{-2}H)\cdot D}.
\end{equation}
Then
\begin{equation}
R=\frac{Y_{N}(\pi ^{+3}H)\cdot Y_{N}(\pi ^{-3}H)}{Y_{N}(\pi ^{-3}He)\cdot
Y_{N}(\pi ^{+3}He)}.
\end{equation}
All normalization quantities not related to $N_{^{3}H}$ and
$N_{^{3}He}$ cancel in $R$. In the definitions of $r_{1}$ and
$r_{2}$, it is the {\em ratio} of the $\pi ^{\pm 2}H$ yields and
cross sections that appears.

Finally, we consider the ratios $\rho ^{+}$ and $\rho ^{-}$. Since the same charge of
pion appears in both the numerator and denominator of each of these ratios, the
non-target-related normalization quantities cancel here as well. Then we have
\begin{equation}
\rho ^{+}=\frac{Y_{N}(\pi ^{+3}H)}{Y_{N}(\pi ^{+3}He)}
\end{equation}
and
\begin{equation}
\rho ^{-}=\frac{Y_{N}(\pi ^{-3}H)}{Y_{N}(\pi ^{-3}He)},
\end{equation}
so that
\begin{equation}
R=r_{1}\cdot r_{2}=\rho ^{+}\cdot\rho ^{-}.
\end{equation}

\section{Ratios}
\label{sec:ratio}

Values for all of the ratios measured in this experiment are given in Table~
\ref{tab:values}. Figure~ \ref{fig:r1r2} shows the angular distributions for the simple
charge-symmetric ratios $r_{1}$, $r_{2}$, and the superratio $R$ at 142, 180, 220, and
256~MeV. We see that $r_{1}$ is flat and structureless in the forward hemisphere but
rises in the backward hemisphere where it remains high to $180^{\circ }$. The ratio
$r_{2}$ shows structure at about $80^{\circ }$ in the forward hemisphere, approaches
1.0 in the backward hemisphere and stays there to $180^{\circ }$. Most of the structure
in R is therefore due to $r_{2}$.
% === PSFIG 2 ======================================
%\psdraft % don't display ps figures

\begin{figure}[htbp]
\centering{
\includegraphics[height=0.80\columnwidth, angle=90]{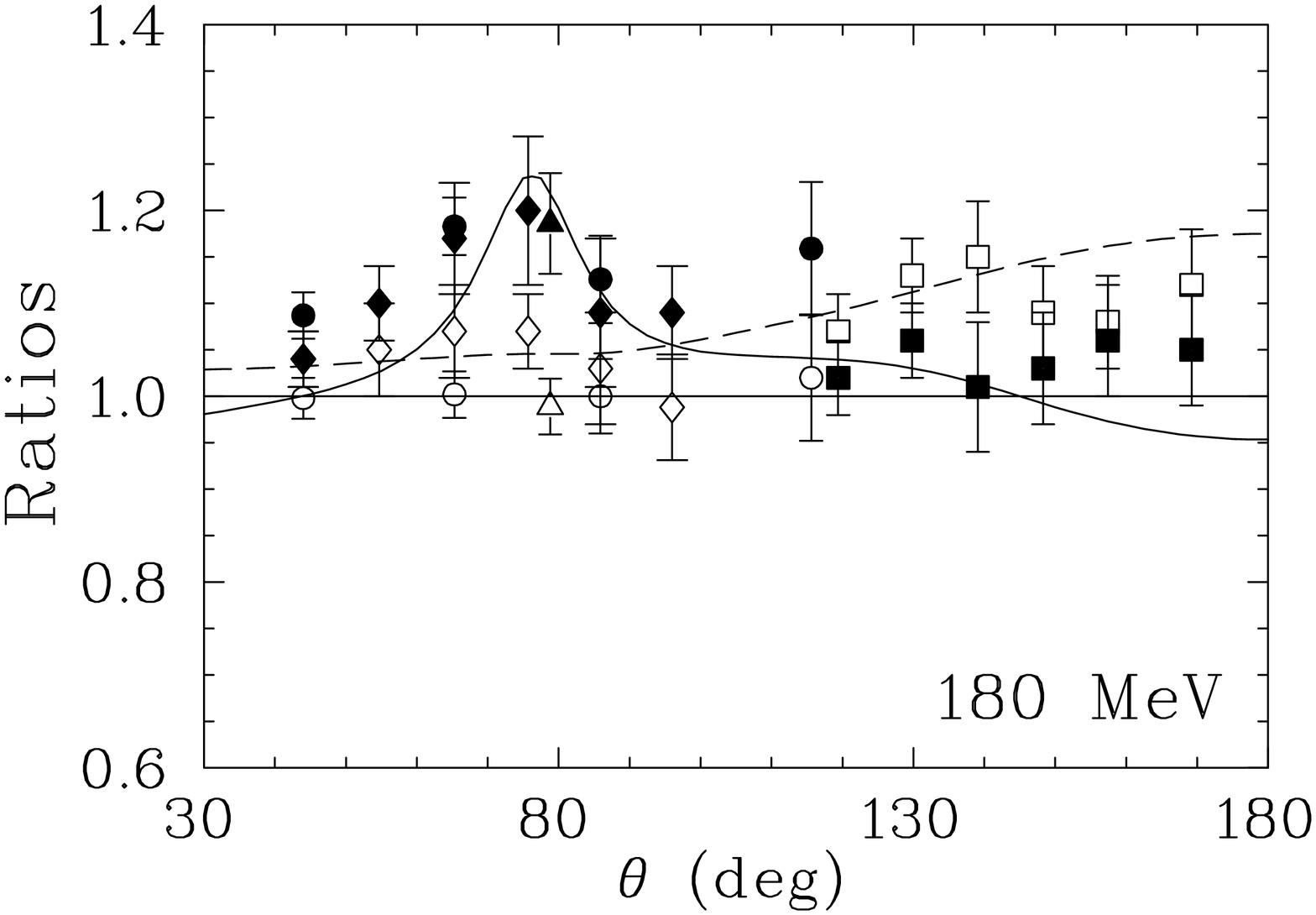}

\vspace{1.5ex}

\includegraphics[height=0.80\columnwidth, angle=90]{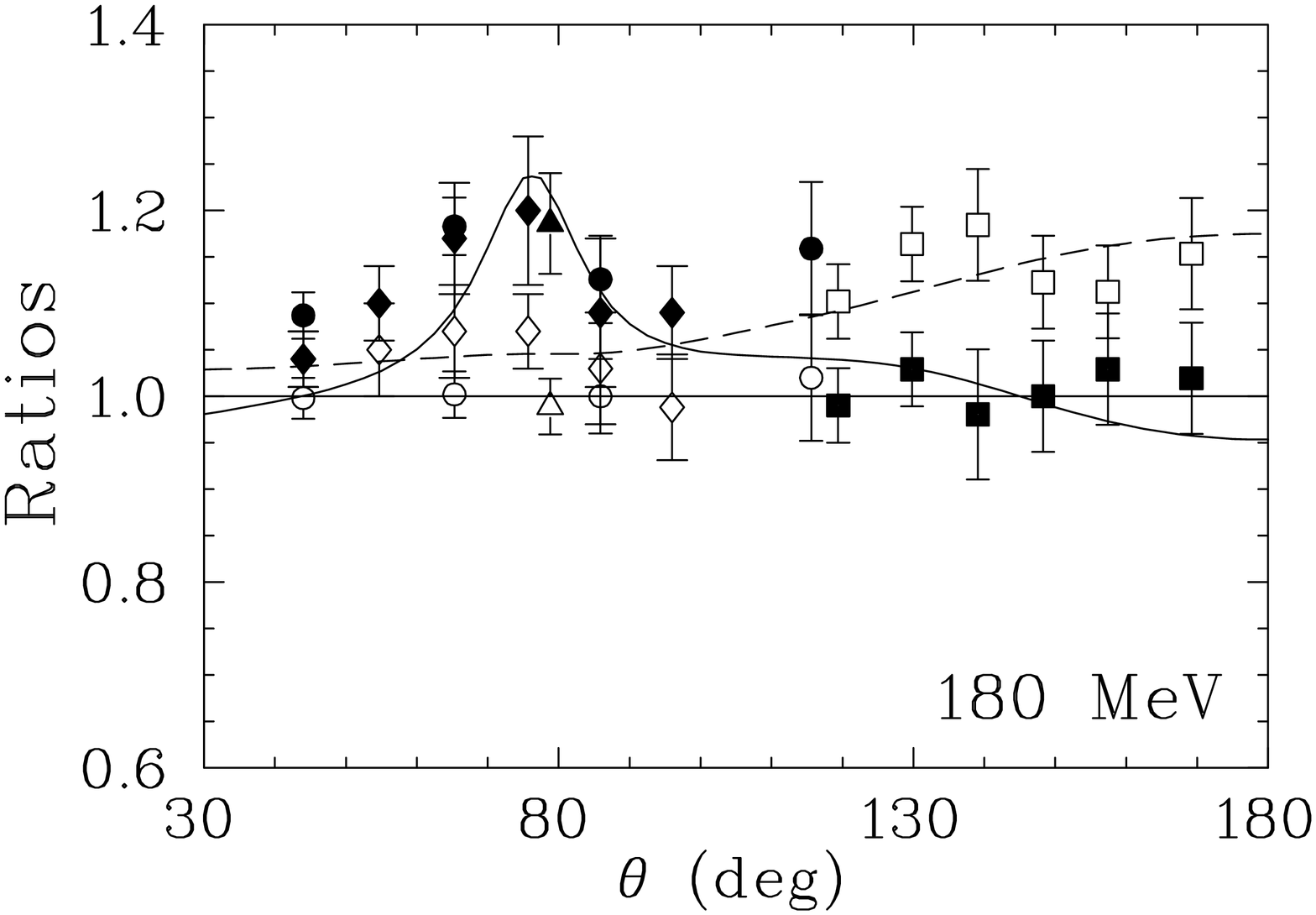}
}
\caption{The simple ratios $r_1$ (open symbols) and $r_2$
         (filled symbols) are compared to the calculations
         of Ref.~\protect\cite{SK02} for $r_1$ (dashed line)
         and $r_2$ (solid line). The qualitative behavior of
         the simple ratios, including the crossover,
         were predicted by an optical-model calculation by
         Gibbs and Gibson~\protect\cite{Gib91} before the
         back-angle data were taken. In the top figure, we used
         the assumption that the yield for {$\pi^+{d} / \pi^-{d}$}
         is equal to 1.00 as is indicated by the SAID fit to the
         {$\pi -d$} data~\protect\cite{said}. In the bottom
         figure, we used the results of Smith \textit{et al.}~
         \protect\cite{Smi88} which give a value of 1.03 for
         this ratio at back angles. A slight improvement of
         the {$\chi^2$} is obtained using the value 1.03. The
         simple ratios cross in the region where double and
         multiple scattering begin to become more important in
         the scattering process.} \label{fig:r1r2cal}
\end{figure}

% === PSFIG 3 ======================================
%\psdraft % don't display ps figures

\begin{figure}[htbp]
\centering{
\includegraphics[height=0.90\columnwidth, angle=90]{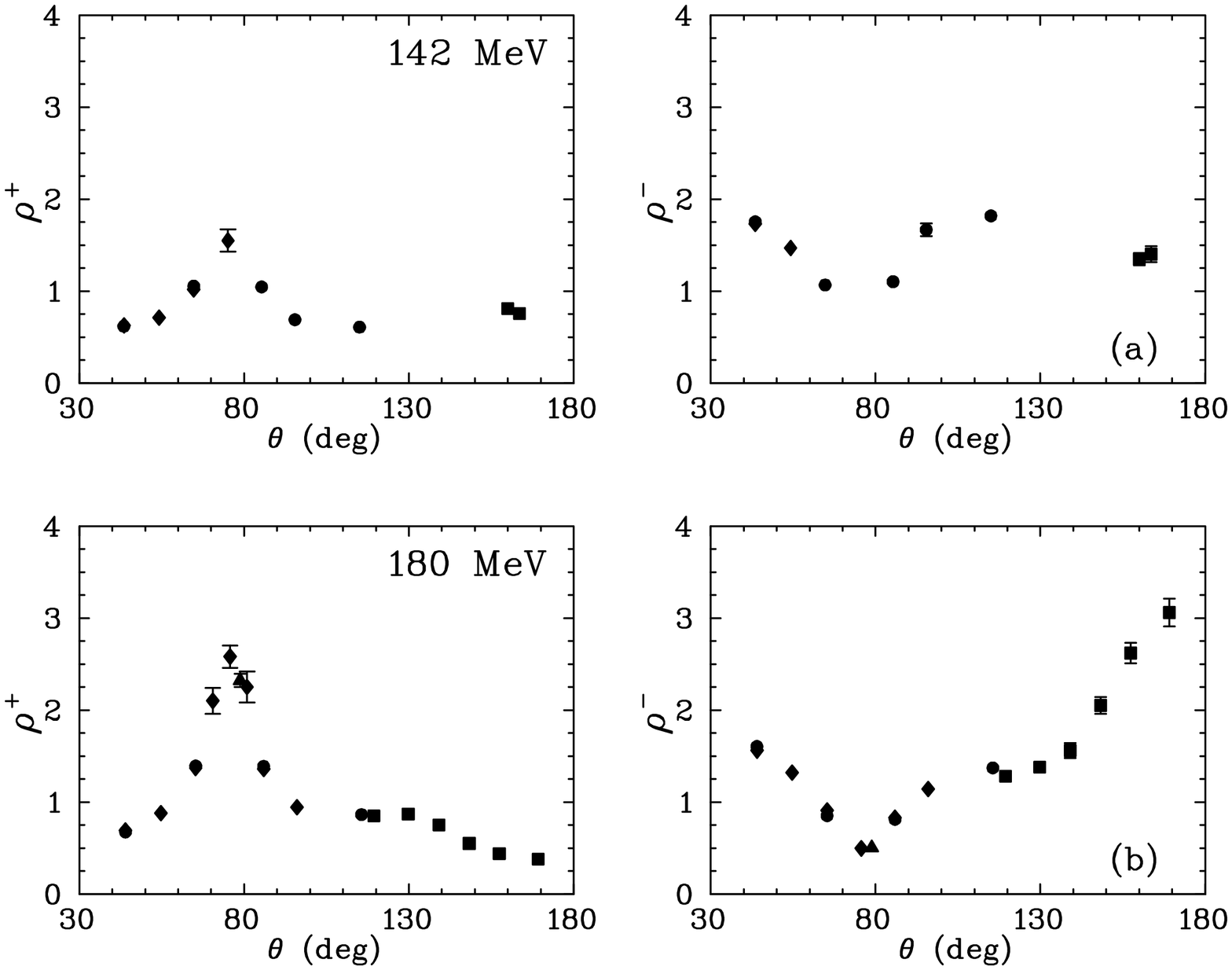}

\vspace{1.5ex}

\includegraphics[height=0.90\columnwidth, angle=90]{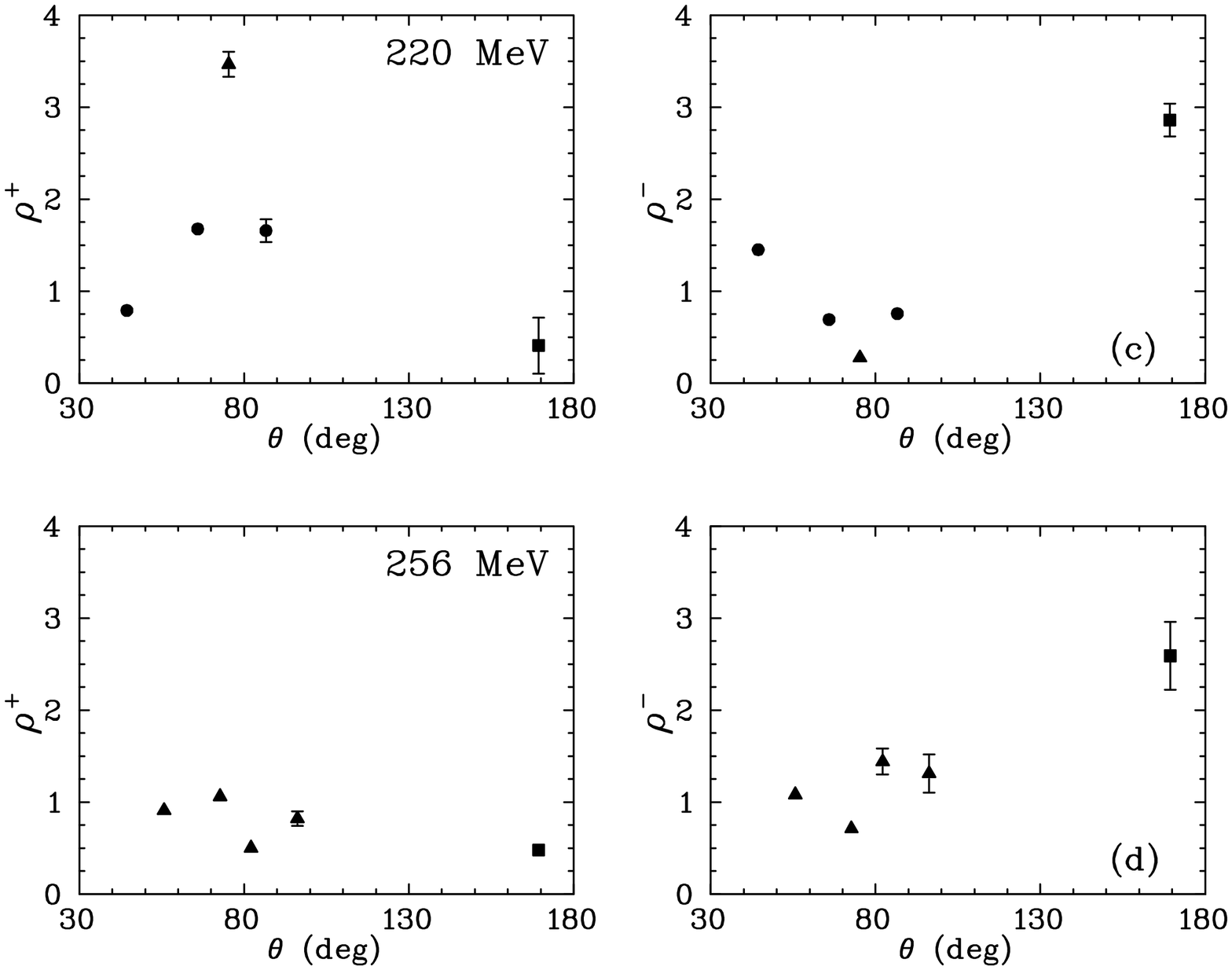}
}
\caption{The ratios $\rho ^+$ and $\rho ^-$
         for (a) T$_{\pi}$ = 142~MeV, (b) 180~MeV,
         (c) 220~MeV, and (d) 256~MeV. The notation
         for the experimental data is the same as for
         Fig.~\ref{fig:r1r2}. } \label{fig:rprm}
\end{figure}

% === PSFIG 4 ======================================
%\psdraft % don't display ps figures

\begin{figure}[htbp]
\centering{
\includegraphics[height=0.9\columnwidth, angle=90]{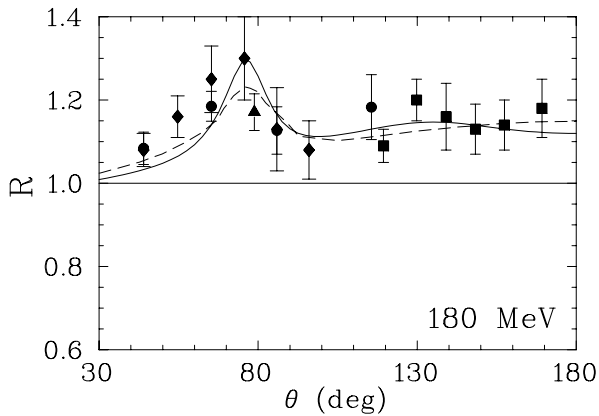}
}
\caption{The Superratio $R$ at T$_{\pi}$ = 180~MeV is
         compared to calculations by
         Kudryavtsev \textit{et al.}~\protect\cite{SK02} (solid
         curve) and Gibbs and Gibson~\protect\cite{Gib91} (dashed
         curve). The notation for the experimental data is the
         same as in Fig.~\ref{fig:r1r2}.}
\label{fig:superratio}
\end{figure}

We note that, aside from the region near $80^\circ$ in the $\pi-^{3}\mathrm{A}$
center-of-mass kinematics, which corresponds to {$90^{\circ }$} in the $\pi-N$ center
of mass where the non-spin-flip scattering amplitudes have zeros, the charge-symmetric
scattering ratios do not have any sharp features. Indeed, in the backward hemisphere
they are fairly flat and quite smooth. This is not surprising, since the $\pi $-nucleon
amplitudes are smooth and have no zeros in this region. The form factors for ${^{3}}H$
and ${^{3}}He$, as measured with electron scattering, are smooth as well. Finally, the
interactions in the numerator and denominator for each ratio are approximately the
same: primarily odd-nucleon in $r_{1}$, and primarily even-nucleon in $r_{2}$. $R$ is
the smooth product of two smooth functions.

Of more interest is the general trend of the ratios as we progress from the forward to
the backward hemispheres, see Fig.~\ref{fig:r1r2cal}. The crossover of the two ratios,
and the fact that $r_{1}$ is significantly different from unity but $r_{2}$ is
consistent with unity is difficult to explain quantitatively. It is interesting to note
that the qualitative behavior of the simple ratios, including the crossover, were
predicted by an optical-model calculation by Gibbs and Gibson~\protect\cite{Gib91}
before the back-angle data were taken.

One might speculate that as one passes to the backward hemisphere, with the diminishing
importance of the single-scattering process, the double-scattering process provides a
mechanism for spin-flip amplitudes to contribute in paired-nucleon scattering without
violating the Pauli exclusion principle. In the top plot of Fig.~\ref{fig:r1r2cal}, we
assume that $d\sigma (\pi ^{+2}H)=d\sigma (\pi ^{-2}H)$. In the bottom plot, we assume
the 1.5\% asymmetry suggested by Smith \textit{et al.}~ \cite{Smi88} for $\theta\ge
90^{\circ}$, which means correcting the value of $D$ by 3\%. The effect is to
\textit{increase} the difference between $r_{1}$\ and $r_{2}$\ in the backward
hemisphere.

Figure~\ref{fig:rprm} shows results for $\rho ^{+}$ \ and $\rho ^{-}$ . These ratios
have very small error bars, owing to the cancellation of normalization quantities
mentioned above. The steep rise of $\rho ^{-}$ and the fall below one of $\rho ^{+}$
are indications that there is a steep rise in the 180-MeV even-nucleon-dominated cross
sections at large angles~ \cite{Mat95}. However, the error bars for these ratios are
much smaller than those for the cross sections. These ratios are the most sensitive
test of the large-angle scattering calculations; in particular, these test the form
factors used in the calculation.

Figure~\ref{fig:superratio} shows $R$\ at 180~MeV. The shape is derived from the two
simple ratios, the bump at $80^{\circ }$ corresponding to the bump in $r_{2}$ and the
steady rise in the backward hemisphere to the rise in $r_{1}$. Figure~
\ref{fig:superratio} also shows the calculations by Kudryavtsev \textit{et
al}~\cite{SK02} and by Gibbs and Gibson~\cite{Gib91}. In the latter the shape of the
Superratio was used to extract differences in the odd and even nucleon radii more
precisely than is possible from existing electron scattering data. Gibbs' and Gibson's
optical-model calculation shows a strong dependence on two parameters, the difference
between the {\em rms } neutron radius in tritium and the {\em rms} proton radius in
${^{3}}He$ (that is, the difference in the even-nucleon radii), and the difference
between the {\em rms} proton radius in ${^{3}}H$ and the {\em rms} neutron radius in
${^{3}}He$. In the approach of Ref.~\cite{SK02}, a successful description of $R$ at
T$_{\pi}$ = 180~MeV is based not only on the difference in the wave functions of
${^{3}}H$ and ${^{3}}He$, but also on the $\Delta_{33}$ mass splitting as well as on
the inclusion of the double-scattering interaction of the pion with nucleons of the
target nuclei. Both models account for the role reversal of the $r_1$ and $r_2$ at back
angles for the $T_\pi$ = 180~MeV. However, Kudryavtsev \textit{et al.} fitted $r_1$ and
$r_2$ to determine R, while Gibbs and Gibson fitted their calculations to R and
inferred $r_1$ and $r_2$ prior to the time that the data were obtained.

Although the leading terms included in the calculation of Kudryavtsev \textit{et al.}
reproduce the main features of the back-angle ratios at energies of 180 MeV and above,
yet another factor which might contribute to this remarkable role reversal of $r_1$ and
$r_2$ at back angles is a two-step process consisting of the formation of a virtual
delta by the interaction of the incident pion with a nucleon, followed by the
scattering of this delta on the remaining correlated nucleon pair. This process is
clearly overshadowed by single scattering at angles near the non-spin-flip dip, but,
like any two-step process, could become important as the momentum transfer increases.

For $r_1$, the dominant channels for delta formation would be $\pi ^{+} + p \rightarrow
\Delta ^{++}$ in the numerator and $\pi ^{-} + n \rightarrow \Delta ^{-}$ in the
denominator, whereas the reverse would be the case for $r_2$. [For $r_1$ the correlated
pairs would be (nn) in the numerator and (pp) in the denominator, and for $r_2$ they
would be (np) pairs in both numerator and denominator.] Since the width of the $\Delta
^{++}$ is somewhat smaller than that of the $\Delta^{-}$ (about 5 MeV out of a total
width of about 120 MeV Ref.~\cite{pdg}), the lifetime of the former would be longer
than that of the latter, and likewise its interaction cross section with the (NN) pair
would be larger. Thus, at back angles, where this effect might become increasingly
important, it would have the effect of increasing $r_1$ and decreasing $r_2$, which
corresponds to their observed behavior.

% === PSFIG 5 ======================================
%\psdraft % don't display ps figures

\begin{figure}[htbp]
\centering{
\includegraphics[height=0.80\columnwidth, angle=90]{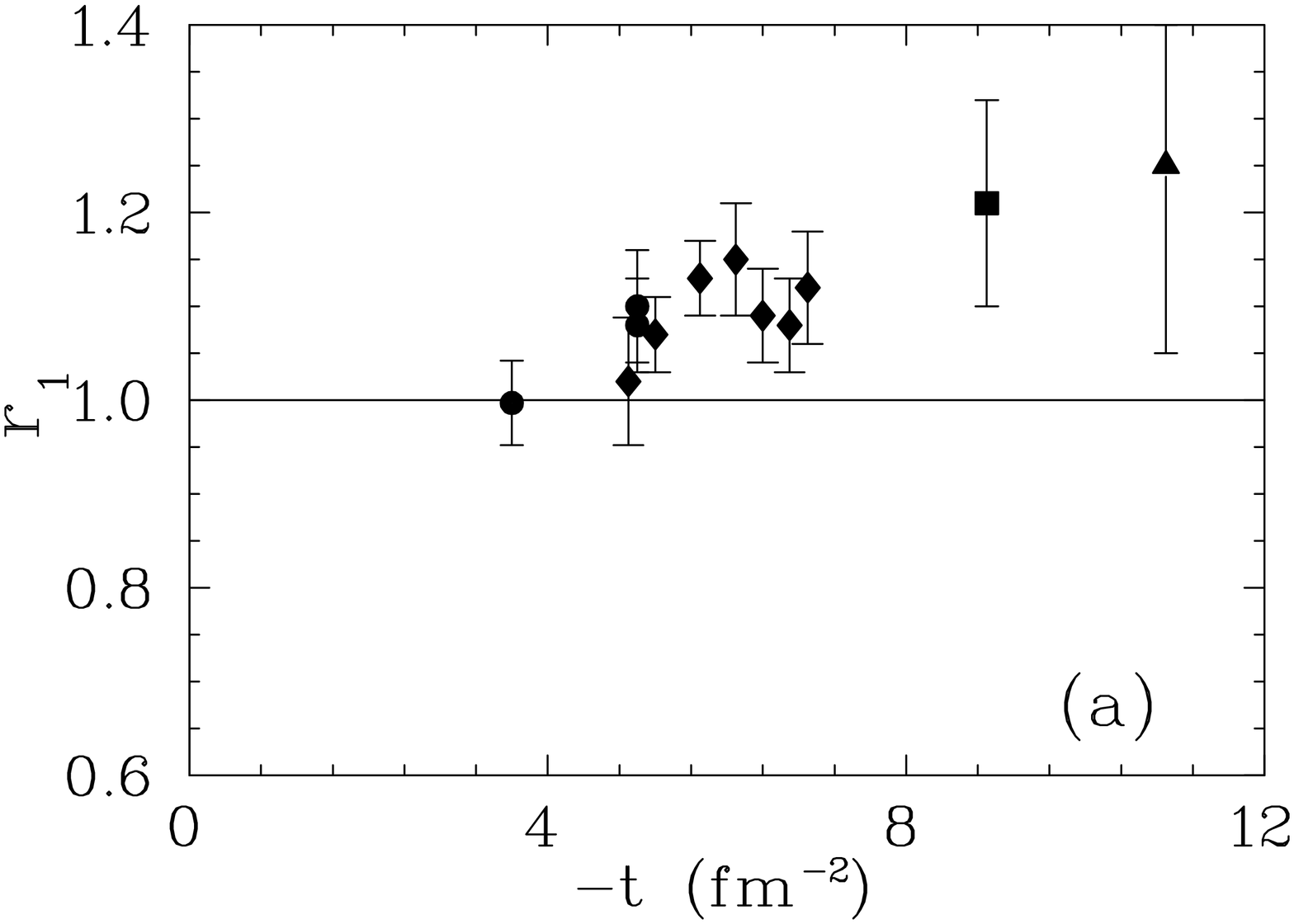}

\vspace{1.5ex}

\includegraphics[height=0.80\columnwidth, angle=90]{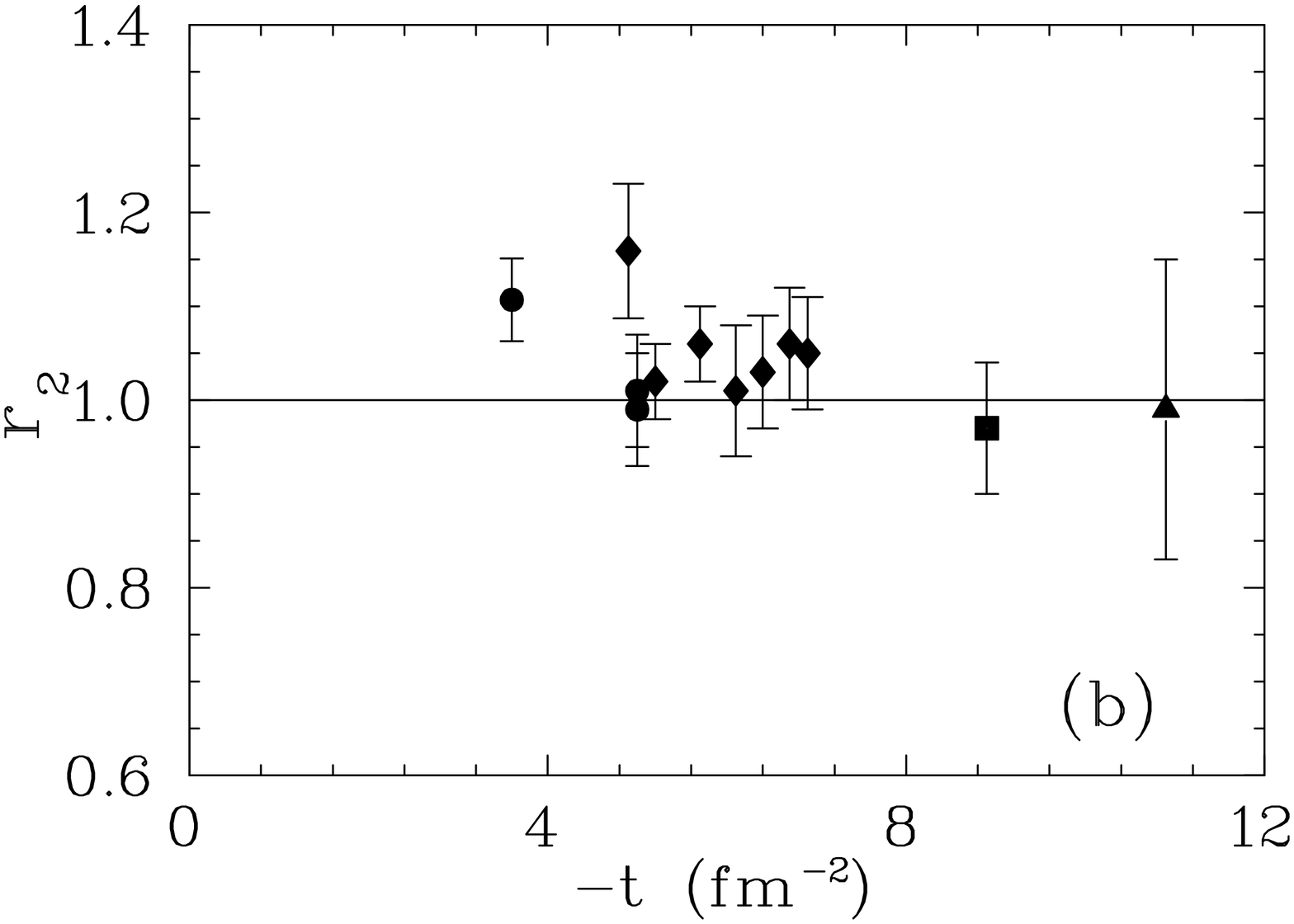}

\vspace{1.5ex}

\includegraphics[height=0.80\columnwidth, angle=90]{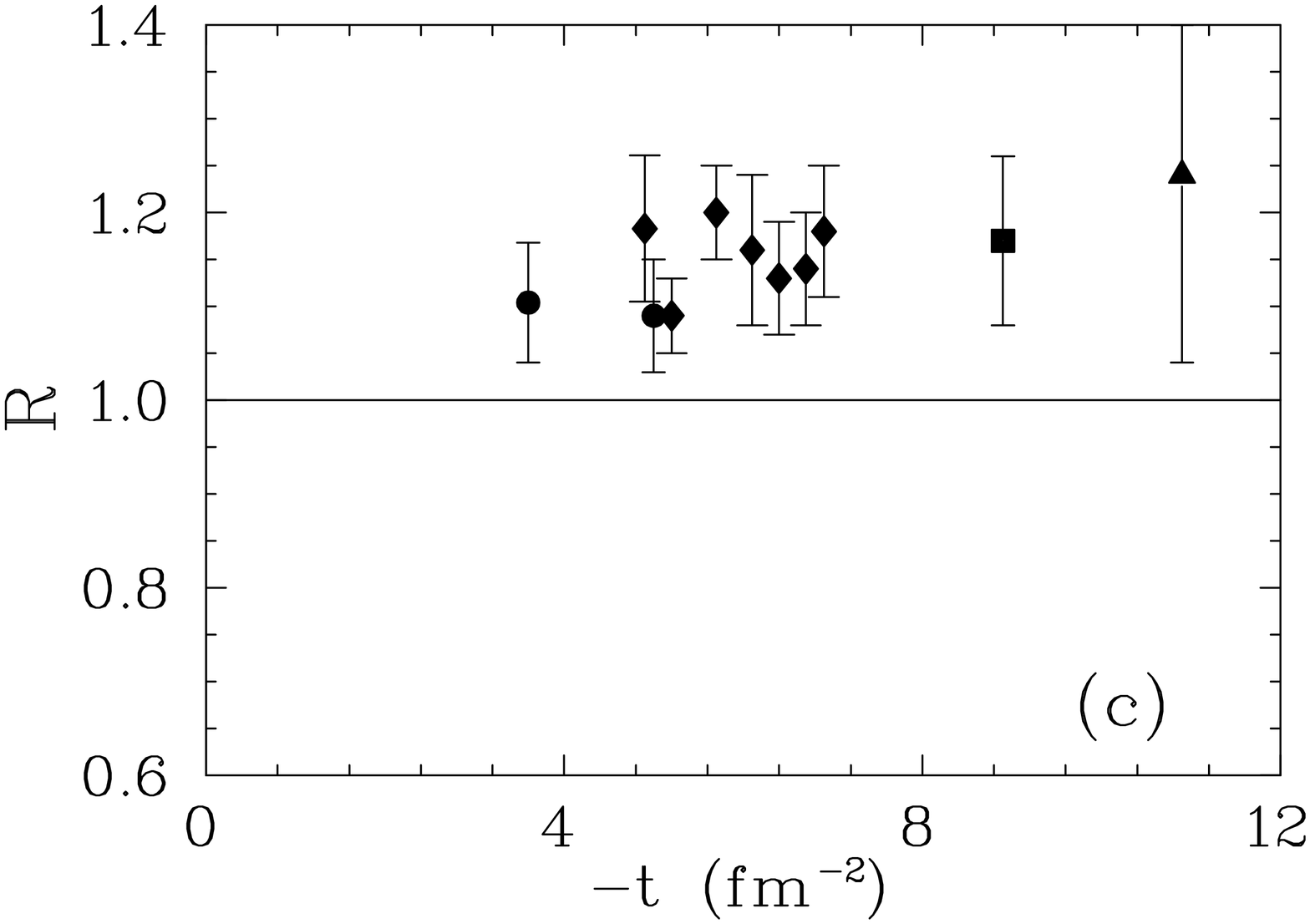}
}
\caption{The excitation function for the ratios (a) $r_{1}$,
         (b) $r_{2}$, and (c) $R$ at back angles ($\ge 100^{\circ}$)
         are shown versus the four-momentum transfer squared $-t$.
         Experimental data are from~\protect\cite{Nef90,Pil91,KSD96}
         and present experiment [142~MeV data are shown by circles,
         180~MeV data by diamonds, 220~MeV data by squares, and
         256~MeV data by triangles].
} \label{fig:r1r2back}
\end{figure}

% === PSFIG 6 ======================================
%\psdraft % don't display ps figures

\begin{figure}[htbp]
\centering{
\includegraphics[height=0.80\columnwidth, angle=90]{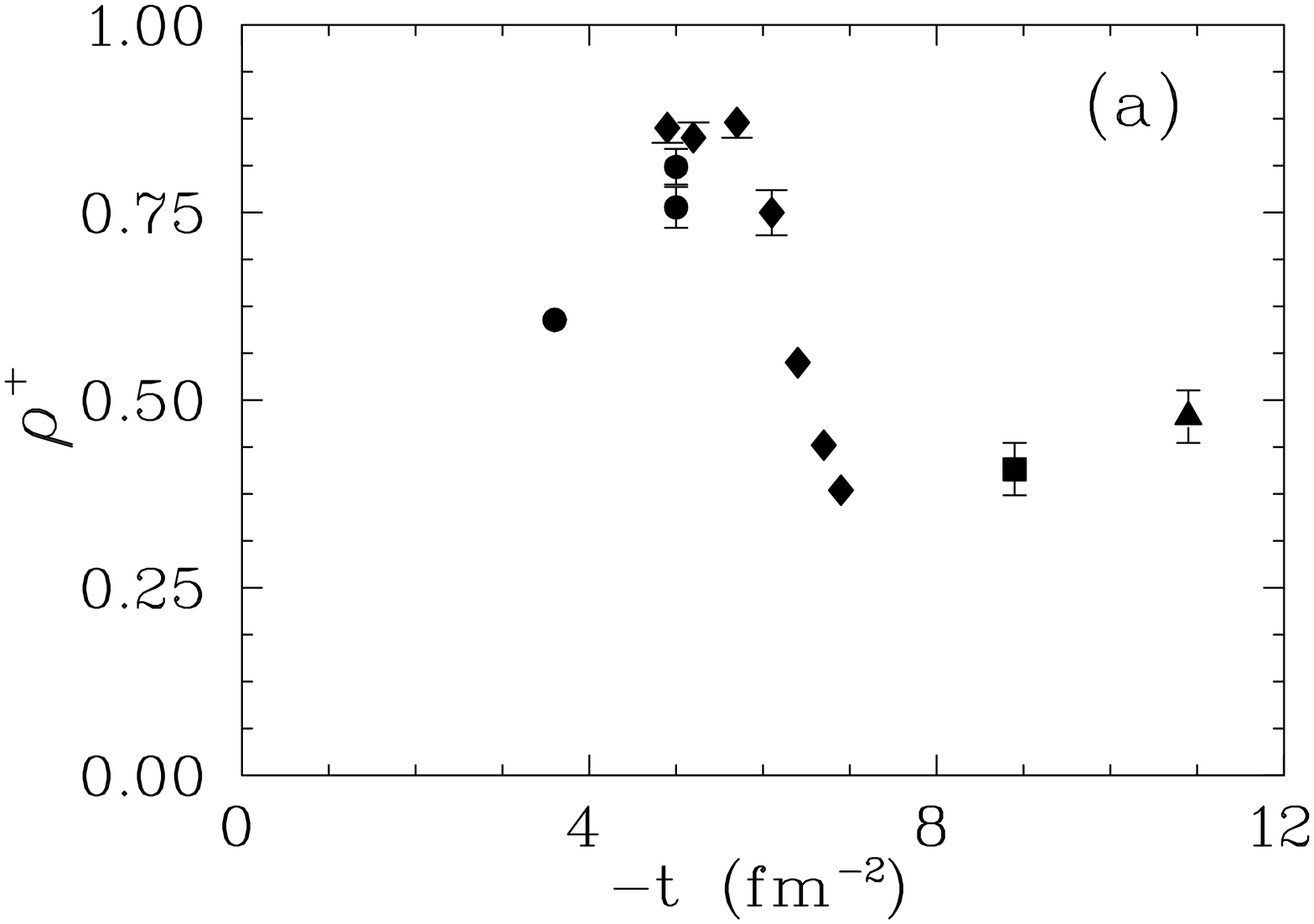}

\vspace{1.5ex}

\includegraphics[height=0.80\columnwidth, angle=90]{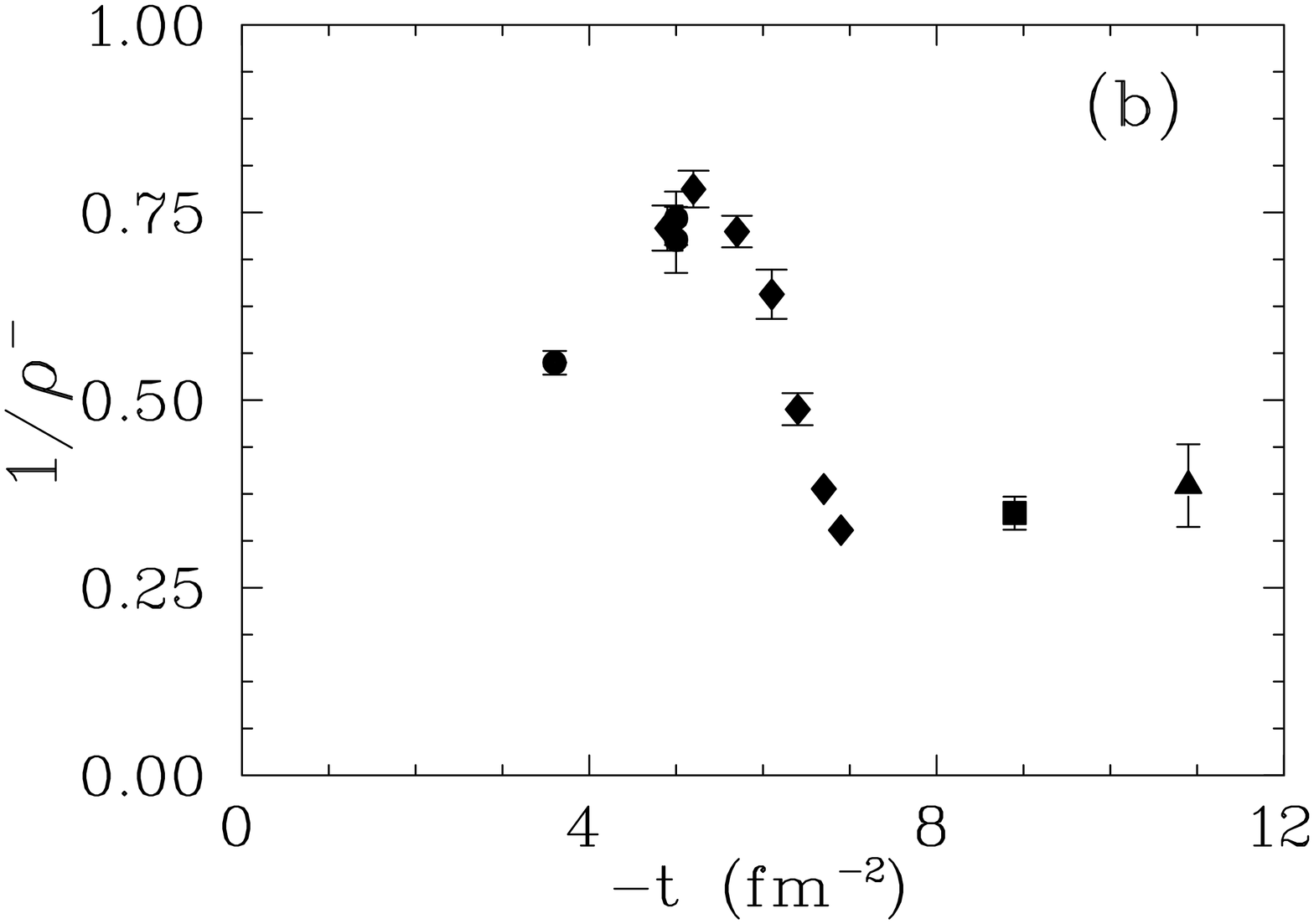}

}
\caption{The excitation function for the ratios (a) $\rho^+$
         and (b) $1/\rho^-$ at back angles ($\ge 100^{\circ}$)
         are shown versus the four-momentum transfer
         squared $-t$. The notation for the experimental
         data is the same as in Fig.~\ref{fig:r1r2back}.
} \label{fig:rho1rho2back}
\end{figure}

\section{Dependence on Momentum Transfer}
\label{sec:90deg}

Figure~\ref{fig:r1r2back} shows $r_{1}$, $r_{2}$, and $R$ at the large scattering
angles ($\ge 100^{\circ}$) for all pion energies plotted versus the four-momentum
transfer squared. In Fig.~5a, $r_{1}$\ increases steadily, while in Fig~5b, $r_{2}$
decreases slightly with -t. $R$ displays a very slight increase with the four-momentum
transfer squared, as shown in Fig.~5c. From Fig.~\ref{fig:r1r2back} one can conclude
that, since $r_{1}$, $r_{2}$, and $R$, coming from different energy sets, fall on top
of each other and follow the same general trend, these ratios are primarily functions
of $-t$.

The behavior of $\rho ^{+}$ \ and $\rho ^{-}$ at the largest scattering angles versus
$T_{\pi}$ can be analyzed with the help of Table 1. As is seen in the table, $\rho
^{+}$ decreases sharply from 142 to 180~MeV, then rises slightly through 220 and
256~MeV. $\rho ^{-}$ shows the opposite behavior, with a maximum at 180~MeV. The
excitation functions for $\rho ^{+}$ \ and $1/\rho ^{-}$ for backward angles versus
$-t$ are shown in Fig.~ \ref{fig:rho1rho2back}. As is the case for $r_1$, $r_2$, and R,
the agreement of the overlaid data at different energies is also quite good.

We note that in the forward hemisphere, a model that assumes single $\pi-N$ scattering
explains the behavior of elastic $\pi-^{3}\mathrm{A}$ scattering quite nicely, and the
reaction is completely defined by $\pi-^{3}\mathrm{A}$ kinematics. However, for angles
greater than $100^\circ$, kinematic considerations force us to consider two-step
processes, especially double scattering to explain $\pi-^{3}\mathrm{A}$ elastic
scattering and all of the ratios are seen to be smooth functions of the momentum
transfer.

\section{Conclusions}
\label{sec:concl}

Several new measurements of the ratios $r_{1}$, $r_{2}$, and $R$ at energies $T_{\pi}$
= 142, 180, 220, and 256~MeV at backward scattering angles are presented. These data
complete our previous data sets at smaller angles and the same energies
~\cite{Nef90,Pil91,WJB93,KSD96}. Where data sets overlap they are consistent with each
other. At $T_{\pi}=180~MeV$, the charge-symmetric ratios $r_{1}$, $r_{2}$, and $R$ are
smooth functions of the scattering angle in the backward hemisphere. The ratios $r_{1}$
and $r_{2}$ cross each other at around $120^{\circ}$; $r_{1}$ becomes significantly
different from 1.00 at backward angles while $r_2$, which had been greater than 1.00 at
forward angles, approaches unity. Deviation of the ratios $r_{1}$, $r_{2}$, and $R$
from unity gives evidence for charge-symmetry-violation effects in these reactions. All
three ratios, at energies of 180, 220, and 256~MeV, are well described by the
theoretical approach of Ref.~\cite{SK02}. Additionally, the ratios at 180 MeV match
well with the results of a previous calculation~\cite{Gib91} that determines the value
of the difference between the even and odd radii of ${^{3}}He$ and ${^{3}}H$. It also
has been shown that all three ratios $r_{1}$, $r_{2}$, and $R$ (as well as $\rho^+$ and
$\rho^-$) at large scattering angles ($\ge 100^{\circ}$) and for all of the incident
pion energies studied can be described as a function of only the four-momentum transfer
squared $-t$ in the region where two-step processes such as double scattering dominate
the $\pi-^{3}\mathrm{A}$ elastic scattering.\\

\section{Acknowledgments}

The authors acknowledge the support of the US National Science Foundation, the US
Department of Energy, and the GW Research Enhancement Fund. We wish to thank
R.~Boudrie, C.~Morris, and S.~Dragic for their assistance during the running of the
experiment and A.~E.~Kudryavtsev and I.~I.~Strakovsky for a careful reading of the
manuscript and for fruitful discussions.

% *** References ***
%*************************************************************************

\end{document}